\documentclass{aa}
\usepackage{psfig}

\def\ltsima{$\; \buildrel < \over \sim \;$}
\def\lsim{\lower.5ex\hbox{\ltsima}}
\def\gtsima{$\; \buildrel > \over \sim \;$}
\def\gsim{\lower.5ex\hbox{\gtsima}}

\begin{document}

\title{M-type giants as optical counterparts of X--ray sources \\
4U 1700+24 and 4U 1954+319\thanks{Partly based on observations 
collected at the Astronomical Observatory of Bologna in Loiano, 
Italy.}}

\author{N. Masetti\inst{1},
M. Orlandini\inst{1},
E. Palazzi\inst{1},
L. Amati\inst{1}
and F. Frontera\inst{1,2}
}

\institute{
INAF - Istituto di Astrofisica Spaziale e Fisica Cosmica di Bologna, 
via Gobetti 101, I-40129 Bologna, Italy (formerly IASF/CNR, Bologna)
\and
Dipartimento di Fisica, Universit\`a di Ferrara, via Saragat 1, I-44100
Ferrara, Italy
}

\titlerunning{M-type giants as counterparts of 4U 1700+24 and 4U 1954+319}
\authorrunning{Masetti et al.}

\offprints{N. Masetti, {\tt masetti@iasfbo.inaf.it}}

\date{Received 15 February 2006; Accepted 9 March 2005}

\abstract{We observed with {\it Chandra} two peculiar galactic X--ray
sources, 4U 1700+24 and 4U 1954+319, which are suspected to have a 
M-type giant star as optical counterpart, in order to get an 
high-precision astrometric position for both of them. The peculiarity of 
these sources lies in the fact that these are the only two cases among 
low-mass X--ray binaries (LMXBs), besides the confirmed case of GX 1+4, 
for which the companion can possibly be a M-type giant. 
We found that in both cases the field M-type giant star
is indeed the counterpart of these X--ray sources. We also determined
the distance to 4U 1954+319 to be $\sim$1.7 kpc. This result suggests 
that a number of faint ($L_{\rm X} \sim$ 10$^{32}$--10$^{34}$ erg 
s$^{-1}$) Galactic X--ray sources are `symbiotic X-ray binaries', 
that is, wide-orbit LMXBs composed of a compact object, most likely a 
neutron star, accreting from the wind of a M-type giant.

}

\maketitle

\section{Introduction}

Low-mass X--ray Binaries (LMXBs) are interacting systems composed of an 
accreting compact object and a low-mass (1 M$_\odot$ or less) 
main-sequence or slightly evolved late-type star. However, among the 
more than 150 LMXBs known (Liu et al. 2001) in only one case 
(GX 1+4; Chakrabarty \& Roche 1997) the secondary is a M-type 
giant. Thus, in principle, LMXBs of this kind are extremely rare.
There are 
however two more LMXBs which might have a M-type giant as optical 
companion, but the X--ray position is not known with sufficiently high 
precision to securely confirm (or reject) the association; moreover, 
their optical spectra do not conclusively show the characteristic 
features of accreting binaries (e.g. Balmer and He {\sc ii} emission 
lines and/or a non-stellar continuum blueward of $\sim$5500 \AA). These 
X--ray sources are 4U 1700+24 and 4U 1954+319.

4U 1700+24 has an X--ray flux which is variable up to a factor of 50 
in the 2--10 keV band, and and X--ray spectrum which shows 
a hardening trend with increasing X--ray flux (Dal Fiume et al. 1990; 
Masetti et al. 2002). The source
exhibits huge X--ray aperiodic time variability: the average fractional
variation in the flux during an {\it EXOSAT} observation was $\sim$50\%
on timescales from tens to thousands of seconds (Dal Fiume et al. 1990).

Masetti et al. (2002) presented an X--ray monitoring (performed
by using data from 4 different satellites and spanning
about 15 years) of this source, showing substantial long-term
variability. Its X--ray spectrum is also variable in the long term.
A high-resolution X--ray spectroscopy analysis of 4U 1700+24 during a 
phase of enhanced activity was performed with {\it XMM-Newton} by Tiengo 
et al. (2005).

Garcia et al. (1983), using {\it EXOSAT} data, identified this source in
the optical with a `normal' M-type giant star, HD 154791, located at
$\sim$800 pc from Earth. Masetti et al. (2002) gave a more accurate 
spectral classification (M2--3 III) to this putative optical
companion and, in turn, a new distance estimate ($d$ = 420 pc).
The latter authors moreover explained the absence of the
features typical of accreting systems in the optical spectrum
of HD 154791 as due to the fact that the luminosity of the giant star is
$\sim$200 times larger that any possible optical emission from the
accreting stream, and so the latter is ovewhelmed by the former.

All of the X--ray characteristics described above are interpreted as 
produced by a neutron star (NS) accreting from the wind of the M-type 
giant. 

The 1.5--12 keV {\it RXTE}/ASM light curve\footnote{available at {\tt 
http://xte.mit.edu/ASM\_lc.html}} of 4U 1700+24 suggests the possibility of 
a periodicity of around 400 days, which would be compatible with the
pulsation period of a M-type giant star or with a wide orbit for the
system (Masetti et al. 2002). Indeed, Galloway et al. (2002) found an
optical spectroscopic period of 404 days, fully consistent with the 
above X--ray finding.

However, Morgan \& Garcia (2001), using a {\it ROSAT}/HRI pointing, gave the 
most precise position for 4U 1700+24 available up to now, with a 
90\%-confidence level error radius of 2$\farcs$1.
This localization is inconsistent, at a 3.6-$\sigma$ level, with the 
extremely accurate position (with 90\% error of 14 mas, considering 
also the proper motion uncertainties) of the M giant provided by {\it 
Hipparcos} (Perryman et al. 1997).
This fact, if confirmed, would suggest that the actual counterpart of 4U 
1700+24 is a background object, and that HD 154791 is probably an interloper 
along the line of sight.

\smallskip

The X--ray source 4U 1954+319 was first detected by the {\it Ariel}
satellite at a flux of $\sim$10 mCrab
in the 2--10 keV range (Warwick et al. 1981), but the first
pointed observation was that by {\it EXOSAT} (Cook et al. 1984).
These authors showed that the source has a dramatic flaring behaviour on
timescales of several minutes, with variations in intensity up to a factor
of 10. This variability is still present, as demonstrated by the
1.5--12 keV {\it RXTE}/ASM light curve of this source.

4U 1954+319 has been subsequently observed by {\it Ginga} (Tweedy et al.
1989), confirming the previous {\it EXOSAT} results. 
The spectrum observed by both {\it EXOSAT} and {\it Ginga}, albeit complex
below $\sim$4 keV, 
is adequately described above this energy by the typical X--ray pulsar
spectrum, namely a power law modified at high energy by an exponential
cutoff (White et al. 1983). 
An in-depth X--ray spectroscopic monitoring of 4U 1954+319 will be 
presented in Rigon et al. (in preparation).

Both the spectral shape and the large amplitude flaring activity seen in 
4U 1954+319 are typical of X--ray binary systems with a NS 
and a high mass companion (e.g., White et al. 1995). Anyway, an 
univocal identification of the optical counterpart of this source is still 
lacking. The two best X--ray positions (J2000) for 4U 1954+319 were provided 
by {\it ROSAT} and are: RA = 19$^{\rm h}$ 55$^{\rm m}$ 41$\fs$5, 
Dec = $+$32$^{\circ}$ 05$'$ 46$''$ (Voges et al. 1999) and 
RA = 19$^{\rm h}$
55$^{\rm m}$ 42$\fs$6, Dec = $+$32$^{\circ}$ 06$'$ 07$''$
({\it ROSAT} Team 2000) with 3$\sigma$ error boxes of radius 9$''$ and
11$''$, respectively. These positions are however inconsistent with each
other; also, in the middle of the two error boxes, and positionally 
inconsitent with both, a $V$-band 12$^{\rm th}$ magnitude M-type 
giant star is present (Tweedy et al. 1989). 
Two more possible fainter candidates (Tweedy et al. 1989) do not show any 
significant H$_\alpha$ activity. Therefore the classification of the
companion of 4U 1954+319 as a close M-type giant (Cook et al. 1984) 
or as a distant and reddened Be star (Tweedy et al. 1989) is still not 
clarified.

In order to conclusively answer, for both X--ray objects, the problem of 
the identification of their true optical counterpart, we performed two
short (`snapshot') observations of their fields in X--rays with 
{\it Chandra}.
Thanks to the superb astrometric accuracy afforded by this spacecraft,
we will restrict the error box regions of 4U 1700+24 and
4U 1954+319 by a factor greater than 10 and 200 in size, respectively.
Moreover, for the latter source the uncertainty between the two available
X--ray positions will be solved. For 4U 1700+24 we also analyzed {\it 
XMM-Newton} archival data to further support the {\it Chandra} results.

As we will show in the following, this comparison will allow us to 
definitely confirm beyond any reasonable doubt that HD 154791 is the true 
optical counterpart of 4U 1700+24, and to determine that the real 
counterpart of 4U 1954+319 is the bright M-type giant mentioned above.
The study of this latter object will be complemented by optical 
spectroscopy afforded in order to determine the spectral type and 
distance to this source.

\section{The {\it Chandra} pointings}

\begin{table*}
\caption[]{Log of the {\it Chandra} and {\it XMM-Newton} observation used 
in this paper.}
\begin{center}
\begin{tabular}{cllcc}
\noalign{\smallskip}
\hline
\noalign{\smallskip}
\multicolumn{1}{c}{Spacecraft} & \multicolumn{1}{c}{Object} & 
\multicolumn{1}{c}{Start day} & Start time & On-source \\
 & & & \multicolumn{1}{c}{(UT)} & time (ks) \\
\noalign{\smallskip}
\hline
\noalign{\smallskip}
{\it XMM-Newton} & 4U 1700+24  & 11 Aug 2002 & 15:38:22 & 5.96 \\
{\it XMM-Newton} & 4U 1700+24  & 07 Mar 2003 & 01:08:08 & 5.02 \\
{\it XMM-Newton} & 4U 1700+24  & 09 Mar 2003 & 01:03:49 & 8.97 \\
{\it XMM-Newton} & 4U 1700+24  & 13 Aug 2003 & 15:23:12 & 13.6 \\
{\it Chandra}    & 4U 1700+24  & 23 Apr 2005 & 02:45:03 & 1.15 \\
{\it Chandra}    & 4U 1954+319 & 18 Jan 2006 & 20:24:30 & 1.18 \\
\noalign{\smallskip}
\hline
\noalign{\smallskip}
\end{tabular}
\end{center}
\end{table*}

We observed 4U 1700+24 (obs. ID: 5455) and 4U 1954+319 (obs. ID: 5456) with 
the HRC-I instrument (Murray et al. 2000) onboard {\it Chandra} (Weisskopf 
et al. 2002) for $\sim$1 ks each. 
The log of these observations is reported in Table 1.
CIAO\footnote{available at {\tt http://cxc.harvard.edu/ciao/}}
v3.2.2 and CALDB\footnote{available at {\tt http://cxc.harvard.edu/caldb/}} 
v3.1.0 were used for the data reduction. The aspect-solution 90\% 
confidence level error radius of 0$\farcs$6 was assumed for the positions 
of the X--ray sources determined from these observations.

The position of the X--ray sources in the {\it Chandra} images was obtained 
using the {\tt celldetect} command. In the 4U 1700+24 data set, only a 
single bright source is detected at the center of the 30$'$$\times$30$'$ 
HRC field of view. For the case of 4U 1954+319, besides the target object, 
a faint (S/N $\sim$ 3) source was detected close to the edge of the 
detector. The {\it Chandra} PSF degradation at large offset 
angles\footnote{see the {\it Chandra} Proposers' Observatory Guide: \\
{\tt http://cxc.harvard.edu/proposer/POG/html/}},
together with the faintness of this object, does not allow one to use it 
to further refine the astrometry of this pointing.

Thus, in both cases, we cannot tie the internal {\it Chandra} astrometry 
to that of optical and/or near-infrared catalogues in order to improve the 
positional uncertainty afforded by the HRC-I X--ray data.

\section{{\it XMM-Newton} archival data}

For the sake of completeness, we also retrived four archival {\it 
XMM-Newton} (Jansen et al. 2001) pointings (Obs. IDs: 0155960601, 
0151240201, 0151240301 and 0151240401) performed on 4U 1700+24 with the 
EPIC instrument (Str\"uder et al. 2001) in `small window' mode between 
August 2002 and August 2003 (see Table 1).
With these data we determined the average X--ray position of the source
using the XMM-SAS\footnote{available at {\tt http://xmm.vilspa.esa.es/sas/}} 
v6.5.0 software.

In this case also, due to the choice of readout mode which guaranteed a
field of view not larger than 4$'$$\times$4$'$, the {\it XMM-Newton}
astrometry could not be tied to optical/near-infrared catalogues; thus,
the accuracy of the {\it XMM-Newton} position for 4U 1700+24 is limited
by the typical EPIC angular resolution, that is, 4$''$ (90\% confidence 
level).

\section{Optical observations}

One medium-resolution optical spectrum of the red star in the field 
of 4U 1954+319 was acquired starting at 23:50 UT of 15 July 2004 in 
Loiano (Italy) with the Bologna Astronomical Observatory $1.52$~metre 
``G.D. Cassini'' telescope plus BFOSC. The exposure time was 1 min. 
The Cassini telescope is equipped with a $1300\times1340$ pixels EEV 
CCD. Grism \#4 and a slit width of $2''$ were used, providing a 
3500-8700 \AA~nominal spectral coverage. The use of this setup 
secured a final spectral dispersion of $4.0$~\AA/pix.

The spectrum, after correction for flat-field, bias and cosmic-ray 
rejection, was background subtracted and optimally extracted (Horne 
1986) using IRAF\footnote{IRAF is the Image Reduction and 
Analysis Facility made available to the astronomical community by the 
National Optical Astronomy Observatories, which are operated by AURA, 
Inc., under contract with the U.S. National Science Foundation. It is 
available at {\tt http://iraf.noao.edu/}}. Wavelength calibration 
was performed using He-Ar lamps, while flux calibration was 
accomplished by using the spectrophotometric standard 
BD +25$^{\circ}$ 3941 (Stone 1977). Wavelength calibration 
uncertainty was $\sim$0.5~\AA; this was checked by using the 
positions of background night sky lines.

\section{Results}

\subsection{4U 1700+24}

The {\it Chandra}/HRC and {\it XMM-Newton}/EPIC X--ray astrometric
results obtained for 4U 1700+24 are reported in Fig. 1 and in Table 2, 
together with all of the most accurate localizations available in the 
literature for this source and for the star HD 154791.
As one can see from Fig. 1, the HRC error box is perfectly consistent
with the position of optical star, and inconsistent at a 3.5-$\sigma$
level with the {\it ROSAT}/HRI localization of Morgan \& Garcia (2001).
Moreover, HD 154791 lies within one HRC detector pixel from the PSF peak 
of the {\it Chandra} X--ray image.

The {\it XMM-Newton} data confirm this result, as they also are 
completely consistent with the more accurate {\it Chandra} position, 
whereas the consistency with the {\it ROSAT}/HRI position is 
apparently marginal.

Rescaling the interloper chance probabilities computed in Morgan \& 
Garcia (2001) using the more precise {\it Chandra} error box, we find
that the numbers computed by these authors become smaller by a factor 
of $\sim$40. 

This confirms, beyond any reasonable doubt, that HD 154791 is indeed the
actual optical counterpart of 4U 1700+24, and that this X--ray source is 
indeed a LMXB composed of a compact object (most likely a NS) orbiting 
a M-type giant and located at 420 pc from Earth (Masetti et 
al. 2002; Galloway et al. 2002).

The HRC-I data do not contain spectral information; however, for 4U 
1700+24 we determine an HRC-I 0.1--10 keV count rate of 2.33$\pm$0.06 
counts s$^{-1}$. This corresponds to a 2--10 keV flux of 
$\sim$2$\times$10$^{-10}$ erg cm$^{-2}$ s$^{-1}$ assuming a 
Comptonization spectrum as described in Masetti et al. (2002) for the 
{\it BeppoSAX} data. This value converts into a 2--10 keV luminosity of 
$\sim$4$\times$10$^{33}$ erg s$^{-1}$ using the above distance estimate 
from Masetti et al. (2002). The comparison of this value with the 
long-term monitoring of this source reported in Masetti et al. (2002) 
indicates that 4U 1700+24 was in an intermediate X--ray activity 
state during the {\it Chandra} observation.

\begin{figure}
\psfig{figure=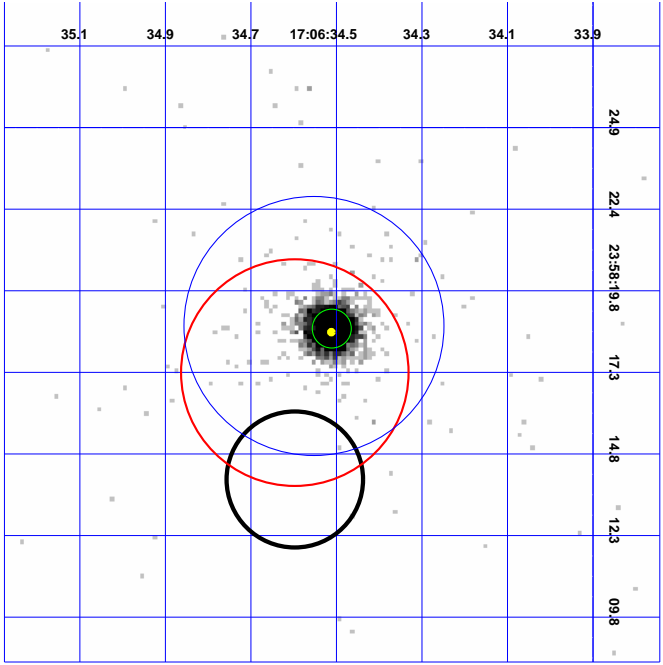,width=8.8cm,angle=0}
\caption[]{{\it Chandra}/HRC image of 4U 1700+24 in the 0.1--10 keV band.
Superimposed to the X--ray image, the following 90\% confidence level 
error circles are drawn: {\it Chandra}/HRC (light circle; radius: 
0$\farcs$6), {\it XMM-Newton}/EPIC (thin dark circle; radius: 4$''$),
{\it ROSAT}/PSPC (medium dark circle; radius: 3$\farcs$5) and
{\it ROSAT}/HRI (thick dark circle; radius: 2$\farcs$1); see Table 2
for details. The {\it Hipparcos} position of the optical star HD 154791 
is indicated by the small light dot. North is at top, East is to the
left. The field size is $\sim$20$''$$\times$20$''$.}
\end{figure}

\begin{table*}
\caption[]{Coordinates (equinox: J2000) and 90\% confidence level errors
of the positions of the X--ray source 4U 1700+24 and of the optical star
HD 154791.}
\begin{center}
\begin{tabular}{llllll}
\noalign{\smallskip}
\hline
\noalign{\smallskip}
\multicolumn{1}{c}{Source} & \multicolumn{1}{c}{RA} & \multicolumn{1}{c}{Dec} 
& \multicolumn{1}{c}{Error} & \multicolumn{1}{c}{Spacecraft} & 
\multicolumn{1}{c}{Reference}\\
 & \multicolumn{1}{c}{(J2000)} & \multicolumn{1}{c}{(J2000)} &  & 
\multicolumn{1}{c}{and instrument} & \\
\noalign{\smallskip}
\hline
\noalign{\smallskip}

HD 154791  & 17$^{\rm h}$ 06$^{\rm m}$ 34$\fs$5183 & 
   +23$^{\circ}$ 58$'$ 18$\farcs$554 & 0$\farcs$014 & {\it Hipparcos} &
   Perryman et al. (1997)\\
4U 1700+24 & 17$^{\rm h}$ 06$^{\rm m}$ 34$\fs$517  & 
   +23$^{\circ}$ 58$'$ 18$\farcs$66 & 0$\farcs$6     & {\it Chandra}/HRC &
   this paper \\
4U 1700+24 & 17$^{\rm h}$ 06$^{\rm m}$ 34$\fs$6    & 
   +23$^{\circ}$ 58$'$ 14$\farcs$0 & 2$\farcs$1      & {\it ROSAT}/HRI &
   Morgan \& Garcia (2001) \\
4U 1700+24 & 17$^{\rm h}$ 06$^{\rm m}$ 34$\fs$6    & 
   +23$^{\circ}$ 58$'$ 17$\farcs$3 & 3$\farcs$5      & {\it ROSAT}/PSPC &
   Morgan \& Garcia (2001) \\
4U 1700+24 & 17$^{\rm h}$ 06$^{\rm m}$ 34$\fs$56   & 
   +23$^{\circ}$ 58$'$ 18$\farcs$7 & 4$''$           & {\it XMM-Newton}/EPIC 
&
   this paper \\

\noalign{\smallskip}
\hline
\noalign{\smallskip}
\end{tabular}
\end{center}
\end{table*}

\subsection{4U 1954+319}

\begin{figure}
\psfig{figure=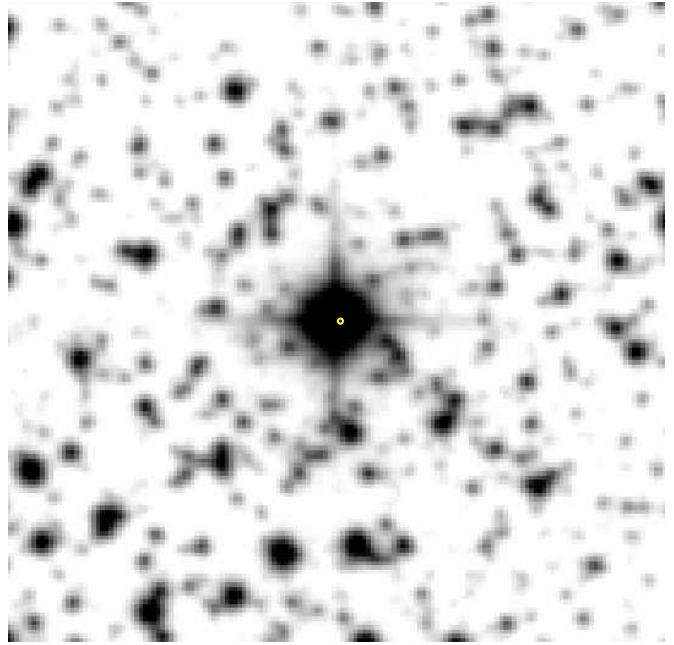,width=8.8cm,angle=0}
\caption[]{DSS-II-Red image of the field of 4U 1954+319 with 
superimposed the 0.1--10 keV band {\it Chandra}/HRC X--ray position
(small circle). As one can see, the HRC position falls on the bright
M-type star in the 4U 1954+319 field. North is at top, East is to the 
left. The field size is $\sim$2$\farcm$5$\times$2$\farcm$5.}
\end{figure}

The position obtained with {\it Chandra}/HRC for 4U 1954+319, overlaid on 
the relevant DSS-II-Red\footnote{available at {\tt 
http://archive.eso.org/dss/dss}} survey image, is shown in Fig. 2. Its 
coordinates, RA = 19$^{\rm h}$ 55$^{\rm m}$ 42$\fs$272, Dec =
+32$^{\circ}$ 05$'$ 48$\farcs$82 (J2000) are consistent within the errors 
with the position of the USNO-A2.0\footnote{The USNO-A2.0 catalogue is 
available at \\ {\tt http://archive.eso.org/skycat/servers/usnoa}} bright 
star U1200\_13816030, that is, the bright red object mentioned in Tweedy 
et al. (1989). This star lies at RA = 19$^{\rm h}$ 55$^{\rm 
m}$ 42$\fs$33, Dec = +32$^{\circ}$ 05$'$ 49$\farcs$1 (with an 
uncertainty of 0$\farcs$25: Assafin et al. 2001; Deutsch 1999), that is, 
0$\farcs$79 from the X--ray position. The two positions are thus
consistent with each other at a 1.8$\sigma$ level, and we can 
confidently state that this star is the optical counterpart of 
4U 1954+319.

\begin{figure*}[th!]
\vspace{-5cm}
\hspace{-1cm}
\psfig{figure=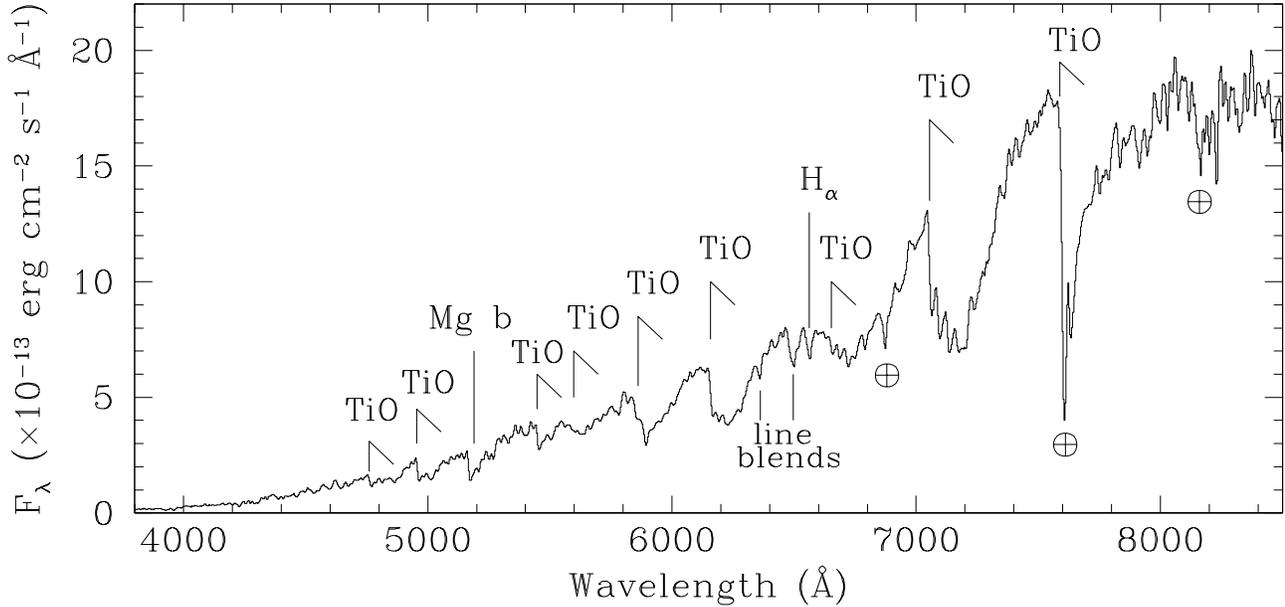,width=20cm,angle=-90}
\vspace{-1.3cm}
\caption[]{3800-8500 \AA~optical spectum of the optical counterpart of
4U 1954+319 obtained with the Loiano 1.52-meter telescope plus BFOSC on 
15 July 2004. The spectrum is typical of a star of type M4-5\,III (see text).
The telluric absorption bands are marked with the symbol $\oplus$.}
\end{figure*}

The inspection of the optical spectrum of this object (reported in 
Fig. 3) clearly shows the typical features of a M-type star (Jaschek 
\& Jaschek 1987): it is dominated by TiO absorption bands and no 
emission features typical of X--ray binaries are apparent. 
The H$_\alpha$ line is also detected 
in absorption. We also detect, among the main spectral features, the 
Mg `b' absorption around 5170 \AA~and two atomic line blends of metal 
intersystem lines of Fe {\sc i}, Ti {\sc i}, Cr {\sc i}, Ba {\sc i}, 
Ca {\sc i}, Mn {\sc i}, Co {\sc i} and Ni {\sc i} located at 6352 
\AA~and 6497 \AA~(see e.g. Turnshek et al. 1985). Telluric absorption 
features are moreover detected at 6870 and 7600 \AA.

Using the Bruzual-Persson-Gunn-Stryker\footnote{available at:\\ {\tt 
ftp://ftp.stsci.edu/cdbs/cdbs1/grid/bpgs/}} (Gunn \& Stryker 1983) and 
Jacoby-Hunter-Christian\footnote{available at:\\ {\tt 
ftp://ftp.stsci.edu/cdbs/cdbs1/grid/jacobi/}} (Jacoby et al. 1984) 
spectroscopy atlases, we then compared the spectrum of star 
U1200\_13816030 with those of several late-type stars. The best 
match is 
obtained with stars HD 110964 (M4\,III) and SAO 62808 (M5\,III). Thus, 
we classify U1200\_13816030 as a star of spectral type M4-5\,III.

Next, from the $R$-band magnitude information extracted from the 
USNO-B1.0 catalogue (Monet et al. 2003) and from the $V-R$ color 
index of a M4\,III star (1.58; Ducati et al. 2001), we determine a 
$V$ magnitude of 10.7 for the counterpart of 4U 1954+319. Assuming 
that a star of this spectral type has an absolute magnitude M$_V$ = 
$-$0.5 (Lang 1992), we obtain a distance $d$ = 1.7 kpc. We however 
stress that this should conservatively be considered as an upper 
limit to the distance to this object, as the effect of the (presently 
unknown) amount of the interstellar absorption along the line of sight 
was not accounted for.

The HRC-I 0.1--10 keV count rate for 4U 1954+319 is 0.014$\pm$0.004 
counts s$^{-1}$, indicating that the source was quite faint at the time 
of the {\it Chandra} observation. This was also confirmed by the {\it 
RXTE}/ASM data for this object. This count rate corresponds to a 2--10 
keV flux of $\sim$1.2$\times$10$^{-12}$ erg cm$^{-2}$ s$^{-1}$ assuming 
a Comptonization spectrum as in Rigon et al. (in preparation). This 
value implies a 2--10 keV luminosity of $\sim$4$\times$10$^{32}$ erg 
s$^{-1}$ using the distance estimate obtained above from the optical 
spectroscopy.

All of the above makes 4U 1954+319 a more distant twin of the 
X--ray binary 4U 
1700+24, both in terms of spectral type of the companion and of X--ray 
luminosity range. This system is therefore most likely hosting a compact 
object accreting from the wind of a M-type giant star showing no spectral 
signatures other than those of its spectral type. The latter 
characteristic is due to the low X--ray luminosity when compared to the 
bolometric luminosity of a M4\,III star ($\sim$2$\times$10$^{36}$ erg 
s$^{-1}$; Lang 1992); so, reprocessing of variable X--ray emission is 
completely lost into the glare of the M-type giant companion. 
Thus, a scenario similar to that described by Masetti et al. (2002) 
for 4U 1700+24 may be applicable to the system 4U 1954+319 also.

\section{Conclusions}

The positional information obtained from the {\it Chandra} observations 
presented in this paper showed that the optical counterparts of X--ray 
sources 4U 1700+24 and 4U 1954+319 are both M-type giant stars to a high 
confidence level. Besides, spectroscopy of the optical counterpart of 4U 
1954+319 allowed us to determine its spectral type and its distance.

The detection of two such systems within the 4$^{\rm th}$ {\it Uhuru} 
catalogue (Forman et al. 1978) indicates that a number of wide 
systems composed of a compact object and a giant star may be present 
in the Galaxy, eventually evolving (as suggested by Gaudenzi \& 
Polcaro 1999) to a tighter configuration similar to that observed in the 
X--ray binary GX 1+4 (Chakrabarty \& Roche 1997).

Thus, in conclusion we suggest that many faint X--ray objects in the 
Galaxy may belong to this subclass of binary systems (which, by 
analogy with symbiotic stars in which a white dwarf accretes from the 
wind of a M-type giant companion, can be dubbed `symbiotic X--ray 
binaries') and that this may be a non-negligible evolution channel 
for LMXBs.

\begin{acknowledgements}

We thank the referee, Hans Ritter, for his useful comments which helped 
us to improve the paper and for suggesting the `symbiotic X--ray binary'
definition for GX 1+4, 4U 1700+24 and 4U 1954+319.
We also thank R. Gualandi for the assistance at the Loiano telescope.
This work has made use of the NASA's Astrophysics Data System and HEASARC
archive, and of the SIMBAD database operated at CDS, Strasbourg, France. 

\end{acknowledgements}

\end{document}